\documentclass[preprint,12pt,times]{elsarticle}

\usepackage[T1]{fontenc}
\usepackage[utf8]{inputenc}
\usepackage{amsmath,amssymb}
\usepackage{graphicx}
\graphicspath{{figures/}}
\usepackage{booktabs}
\usepackage{tabularx}
\usepackage{array}
\usepackage{multicol}
\usepackage{xcolor}
\usepackage{microtype}
\usepackage{xurl}
\usepackage[hidelinks]{hyperref}
\journal{Renewable and Sustainable Energy Reviews}
\biboptions{sort&compress}
\newcommand{\wordcount}{approximately 7,300}

\newcommand{\tabfont}{\footnotesize}
\newcolumntype{Y}{>{\raggedright\arraybackslash}X}

\begin{document}

\begin{frontmatter}

\title{A Critical Review of Grid Integration Challenges and Flexibility Opportunities for Large-Scale Data Centers}

\author[wse]{Bibek Dhakal\corref{cor1}}
\ead{bibedhak@ttu.edu}
\author[reng]{Manohar Chamana}
\author[reng]{Biswambhar Gayen}
\author[ece]{Jatendar Kumar}
\author[ece]{S M Shams Rahman}
\author[pnnl]{Aaqib Peerzada}
\author[pnnl]{Ki Yeob Lee}
\author[nwi]{Olatunji M. Adeyanju}
\author[pnnl]{Bhaskar Mitra}
\author[ece]{Stephen Bayne}

\cortext[cor1]{Corresponding author.}

\affiliation[wse]{organization={Wind Science and Engineering Program, National Wind Institute, Texas Tech University},
  city={Lubbock}, postcode={79409}, state={TX}, country={USA}}
\affiliation[reng]{organization={Renewable Energy Program, National Wind Institute, Texas Tech University},
  city={Lubbock}, postcode={79409}, state={TX}, country={USA}}
\affiliation[ece]{organization={Department of Electrical and Computer Engineering, Texas Tech University},
  city={Lubbock}, postcode={79409}, state={TX}, country={USA}}
\affiliation[nwi]{organization={National Wind Institute, Texas Tech University},
  city={Lubbock}, postcode={79409}, state={TX}, country={USA}}
\affiliation[pnnl]{organization={Electrical Infrastructure and Buildings Division, Pacific Northwest National Laboratory},
  city={Richland}, postcode={99352}, state={WA}, country={USA}}

\begin{abstract}
The rapid proliferation of cloud computing and artificial intelligence workloads is fundamentally reshaping the magnitude, spatial concentration, and electrical characteristics of data center power demand. Data centers differ from conventional commercial loads because they combine high rack densities, converter-dominated interfaces, thermally coupled cooling, uninterruptible power supplies, programmable workloads, and increasingly complex behind-the-meter generation and storage within a single facility. This review examines their grid integration from utility and sustainable-energy perspectives across resource planning, large-load interconnection, distribution feeders, market operation, real-time dynamics, power quality, and asset management. Particular attention is given to distribution-level issues that receive limited treatment in broader energy reviews: bidirectional operation on renewable-rich feeders, harmonic propagation and resonance, ride-through and reconnection, transformer derating, cable ampacity, neutral-conductor heating, protection, and measurement. Static, composite, physics-based dynamic, calibrated, impedance, electromagnetic transient, and hybrid load models are compared according to study purpose and data requirements, and utility coordination platforms, uninterruptible power supply and battery control, workload flexibility, renewable co-location, cogeneration, fuel cells, thermal storage, and clean firm supply are assessed. The central finding is that requested peak demand and annual energy matching alone cannot establish reliable or sustainable integration. Credible assessment must consider hourly and locational supply, network deliverability, marginal emissions, lifecycle and water impacts, backup obligations, rebound, asset health, and verified operating capability. Nine research priorities are identified, including measurement-driven models, distribution-level performance requirements, updatable model governance, international comparison, and transparent sustainability counterfactuals.
\end{abstract}

\begin{highlights}
\item Critical review of data center grid integration from bulk planning to feeder assets
\item Peak megawatts and annual energy cannot characterize data center grid risk
\item Harmonic, transformer-aging, and cable-ampacity mechanisms are mapped to feeders
\item Load-model classes are matched to study questions, time scales, and data needs
\item Nine research priorities link measurement, model governance, and sustainability
\end{highlights}


\end{frontmatter}

\noindent\textbf{Word count:} \wordcount\ words (abstract, main text, and headings; excluding title, author details, keywords, highlights, this list of abbreviations, table and figure content, declarations, and references).

\section*{List of abbreviations}
\begin{multicols}{2}
\footnotesize\raggedright\setlength{\parindent}{0pt}\setlength{\hangindent}{1.2em}
ADMS: advanced distribution management system\\
AI: artificial intelligence\\
BESS: battery energy storage system\\
CEMS: campus energy management system\\
CHP: combined heat and power\\
DER: distributed energy resource\\
DERMS: distributed energy resource management system\\
EHV: extra-high voltage\\
EMS: energy management system\\
EMT: electromagnetic transient\\
FERC: Federal Energy Regulatory Commission\\
H-OPF: harmonic-constrained optimal power flow\\
HV: high voltage\\
ISO: independent system operator\\
IT: information technology\\
LV: low voltage\\
MV: medium voltage\\
NERC: North American Electric Reliability Corporation\\
OPF: optimal power flow\\
PCC: point of common coupling\\
PDU: power distribution unit\\
PFC: power-factor correction\\
PMU: phasor measurement unit\\
POI: point of interconnection\\
PUE: power usage effectiveness\\
PV: photovoltaic\\
QSTS: quasi-static time series\\
RMS: root mean square\\
RTO: regional transmission organization\\
SCADA: supervisory control and data acquisition\\
THD: total harmonic distortion\\
UPS: uninterruptible power supply\\
VFD: variable-frequency drive\\
ZIP: constant impedance, constant current, constant power (load model)
\end{multicols}

\section{Introduction}
\label{sec:introduction}

Data centers support cloud services, enterprise computing, artificial intelligence (AI), high-performance computing, and digital communication. Their electricity use has become a planning and operating concern for power systems rather than only an internal efficiency issue. Multiple assessments project substantial growth, but their ranges differ because they depend on workload adoption, hardware and cooling efficiency, utilization, project realization, and siting assumptions \cite{ref3,masanet2020recalibrating,mytton2022sources,ref253,shehabi2024report}. Earlier estimates provide useful historical baselines but should not be extrapolated directly to the current AI-driven buildout \cite{koomey2011growth,ref9}. Recent North American Electric Reliability Corporation (NERC) reliability assessments and Federal Energy Regulatory Commission (FERC) staff market reports treat large computational loads as material considerations in resource adequacy, transmission planning, and electricity markets \cite{ref253,ref267}.

Annual energy use and requested peak demand do not, by themselves, describe grid risk. Two facilities with the same peak megawatt rating may have materially different effects. One may have controlled ramps, verified ride-through, timely telemetry, and coordinated storage; another may have voltage-sensitive protection, uncertain converter behavior, or a weak point of interconnection (POI). Data centers are better represented as concentrated, converter-interfaced, thermally coupled loads with high service-continuity requirements and partly programmable demand than as a single static demand block \cite{zhabelova2015data,sun2022dynamic,al2025programmable}.

The literature has developed along several partly disconnected lines. Facility reviews address cooling, power usage effectiveness (PUE), internal energy modeling, and integrated energy systems \cite{ahmed2021review,dayarathna2015data,ref14}. Flexibility and storage reviews examine workload scheduling, uninterruptible power supply (UPS) systems, batteries, and ancillary services \cite{takci2025data,ref44,mohammadi2026grid}. Power-system surveys \cite{chen2025electricity,ginzburg2025technical,viththarachchige2026concerns} and regional forecasts \cite{ref1} examine AI demand, planning, markets, and stability. Less synthesis connects these topics to feeder behavior, measurement-driven load models, power quality, equipment aging, and implementation of emerging reliability requirements.

The scope is the utility interface of large computational loads across grid layers and time scales. This interface is not confined to the transmission system. Although the largest hyperscale campuses typically connect at transmission voltage through dedicated substations, a substantial share of near-term growth is expected to be served from shared, mixed-use distribution systems, and distribution utilities are consequently receiving a growing volume of large-load interconnection requests that range from multi-megawatt edge facilities to campus expansions of tens of megawatts \cite{krishnamoorthy2026role,force2025characteristics}. Studies of data centers report effects on feeder reliability \cite{fan2025reliability}, phase balance \cite{wang2017phase}, voltage control \cite{yan2026switching}, and reconfiguration and hosting capacity \cite{qiu2026characterizing}. Adjacent work on power-electronic loading provides the transferable transformer and asset mechanisms needed to interpret these local effects \cite{mitra2023analyzing,nair2026predictive}.

\begin{figure}[!tp]
\centering
\includegraphics[width=\textwidth,height=0.7\textheight,keepaspectratio]{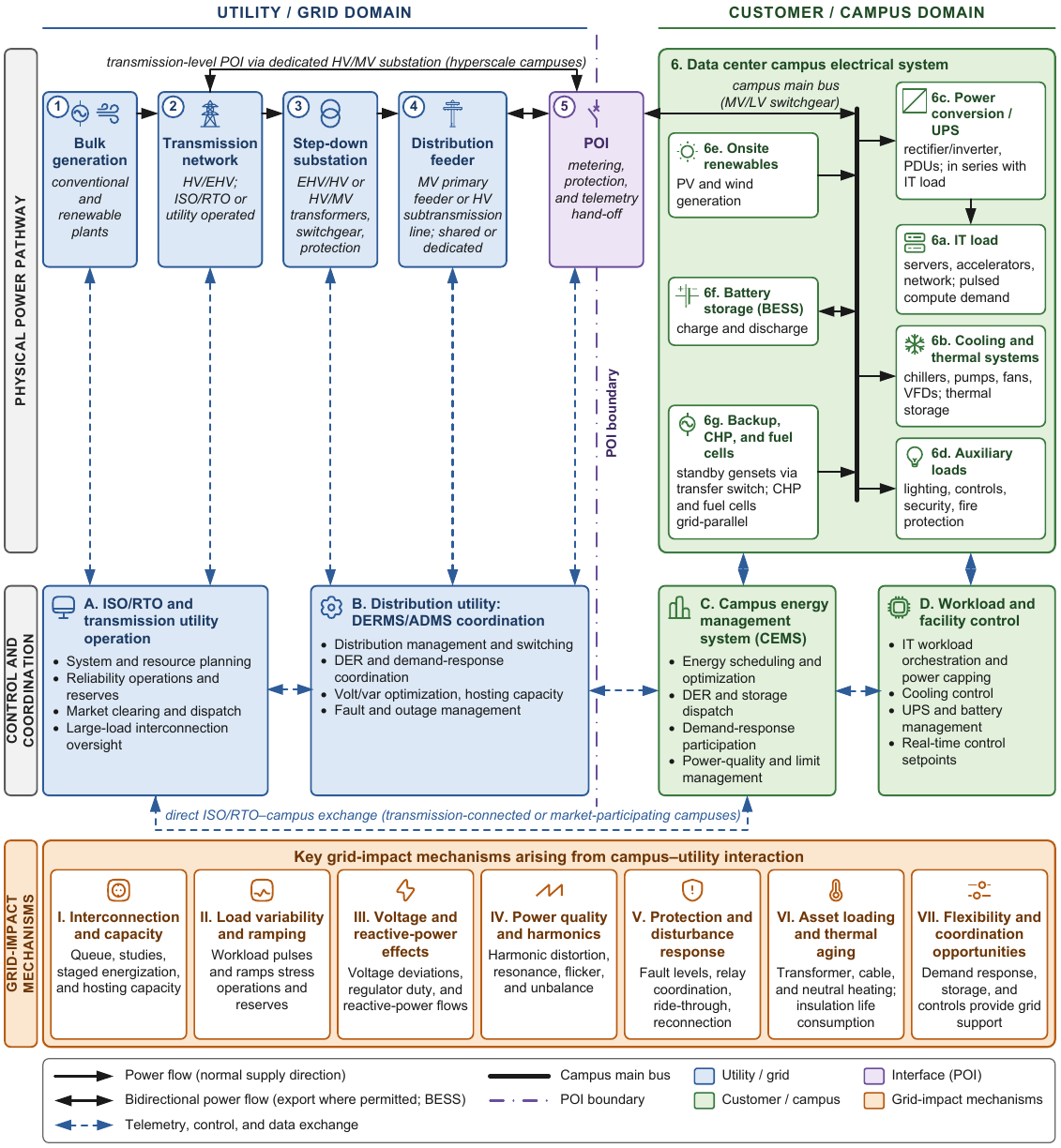}
\caption{Grid-layer and control-domain taxonomy for data center integration. Solid arrows denote physical power exchange, dashed arrows denote telemetry and control, and the vertical boundary marks the point of interconnection (POI). Connection voltage, ownership, and the location of co-located resources vary with facility scale and local topology. The seven grid-impact mechanism groups (interconnection and capacity; load variability and ramping; voltage and reactive-power effects; power quality and harmonics; protection and disturbance response; asset loading and thermal aging; and flexibility and coordination opportunities) arise from the interaction between the campus and the utility network \cite{quint2025practical,krishnamoorthy2026role,force2025characteristics}.}
\label{fig:grid-layer-taxonomy}
\end{figure}

This review develops an integrated grid-layer and time-scale taxonomy that links facility characteristics to planning, interconnection, distribution operation, markets, dynamics, power quality, and asset aging. It compares static, composite, physics-based dynamic, calibrated, converter, electromagnetic transient (EMT), and hybrid load models. It also connects harmonic-rich demand with transformer derating \cite{mitra2023analyzing,faiz2015derating}, cable ampacity \cite{demoulias2007ampacity}, neutral-current accumulation \cite{gruzs1990survey}, and asset management using representative transformer, cable, and computer-load evidence. NERC \cite{force2025characteristics,ref248} and FERC \cite{ref259,ref267} developments are interpreted as a North American implementation case for reliability practice, computational-load modeling, interconnection, and market design. Mitigation pathways are organized around distributed energy resource management systems (DERMS) and advanced distribution management systems (ADMS) \cite{strezoski2023distributed}, UPS and storage control \cite{ross2026using,azizi2026strengthening}, workload scheduling and demand-side measures \cite{ref155}, power-quality design, asset monitoring, and supply co-location. Figures~\ref{fig:grid-layer-taxonomy} and~\ref{fig:timescale-taxonomy} define the organizing framework for the remaining sections.

\section{Review methodology, scope, and positioning}
\label{sec:scope-framework}

\subsection{Scope, source selection, and evidence grading}

This article is a problem-oriented critical review of the grid interface of data centers and other large computational loads. The analysis spans facility operation, resource and transmission planning, interconnection, distribution-system performance, market participation, real-time dynamics, power quality, asset degradation, and sustainable supply. The synthesis asks how data centers differ from conventional large loads, how those differences propagate across grid layers and time scales, which measurements and model classes are appropriate for each study objective, and which supply, control, and flexibility strategies can mitigate reliability and sustainability risks.

Sources were assembled in three groups. Peer-reviewed journal and conference papers were prioritized for technical mechanisms, model development, experiments, and quantified results. Standards, national-laboratory, utility, and policy reports, and official NERC and FERC records were used to describe current engineering practice, reliability concerns, and regulatory implementation. Preprints were included only where they provide the earliest available evidence on rapidly developing topics such as ride-through, reconnection, resonance, and calibrated data center models; they are identified as preprints in the reference list and are treated as emerging rather than established evidence. Direct data-center measurements and models were distinguished from adjacent-mechanism studies of nonlinear computer loads, converters, transformers, cables, protection, and renewable-rich feeders, and the latter were retained only where the physical mechanism is transferable and their limitations could be stated explicitly. The resulting evidence base comprises 134 sources published between 1986 and 2026. Publication status and active regulatory proceedings were checked at the time of writing; because these records can change, their status is reported wherever it affects interpretation.

The review does not assume that one metric or one model can characterize every facility. Annual energy, peak demand, PUE, flexibility, harmonic distortion, and interconnection capacity answer different questions. The analytical framework therefore relates each claimed impact to the relevant grid layer, time scale, data requirement, model fidelity, and implementation constraint, and the evidence grading introduced here is carried through Tables~\ref{tab:asset-degradation}, \ref{tab:model-classes}, and~\ref{tab:impact-solution}, which separate established mechanisms from emerging evidence and modeled potential from verified operating capability. Making these distinctions explicit is necessary because much of the data center grid literature is recent, partly not yet peer reviewed, and changing quickly; without it a reader cannot judge which conclusions rest on field measurement, which on simulation, and which on regulatory text.

\subsection{Positioning relative to existing reviews}

Prior reviews fall into four overlapping groups: energy-estimation studies \cite{masanet2020recalibrating,mytton2022sources}, facility reviews \cite{ahmed2021review,dayarathna2015data,ref14}, flexibility and storage reviews \cite{takci2025data,ref44,mohammadi2026grid}, and broader AI and grid surveys \cite{ahrabi2025ai,chen2025electricity,ginzburg2025technical}. The power-system surveys establish the importance of demand growth, planning, markets, and stability, but a less developed connection remains between those topics and feeder operation, distribution asset degradation, measurement and calibration of computational-load models, power quality, and implementation of reliability requirements \cite{viththarachchige2026concerns}. This review therefore uses the utility interface as its organizing unit, extending from large-load studies and hosting capacity to ride-through, reconnection, converter dynamics, harmonic propagation, equipment derating, and asset aging.

Official reliability and regulatory records provide implementation context rather than substitutes for peer-reviewed technical studies. NERC records address risk characterization, practice gaps, interconnection studies, commissioning, and computational-load modeling \cite{force2025characteristics,ref244,ref248}, and FERC proceedings address transmission interconnection, co-location, forecasting, and market implications \cite{ref259,ref264,ref266}. These materials are treated as a North American case, not as a universal regulatory framework. Table~\ref{tab:existing-reviews} positions the present contribution relative to the major review streams.

\begin{table}[!t]
\centering
\tabfont
\setlength{\tabcolsep}{4pt}
\renewcommand{\arraystretch}{1.1}
\caption{Positioning of the present review relative to existing literature streams}
\label{tab:existing-reviews}
\begin{tabularx}{\textwidth}{Y Y Y >{\raggedright\arraybackslash}p{0.12\textwidth}}
\toprule
\textbf{Literature stream} & \textbf{Representative coverage} & \textbf{Contribution developed here} & \textbf{Key citations} \\
\midrule
Energy estimates and growth & Historical and global data center electricity estimates and uncertainty in projections. & Relates demand growth to grid-layer impacts, interconnection requirements, and model needs. & \cite{ref3,masanet2020recalibrating,mytton2022sources,ref9,shehabi2024report} \\
Facility energy and cooling & Cooling technologies, PUE, energy-consumption models, waste heat, and internal reliability. & Connects internal load composition and controls with grid-facing dynamics, power quality, and distribution assets. & \cite{ahmed2021review,dayarathna2015data,ref14,gnibga2026improving,winick2025reducing} \\
Storage and flexibility & UPS systems, batteries, hybrid storage, ancillary services, workload shifting, and flexible computing. & Separates modeled flexibility from verified capability and places it alongside interconnection, standards, and asset measures. & \cite{takci2025data,ref44,mohammadi2026grid} \\
Broad AI and data center grid reviews & AI infrastructure, demand stages, planning, markets, stability, and broad solution categories. & Adds a distribution-centered treatment of power quality, asset degradation, model governance, and implementation requirements. & \cite{chen2025electricity,ginzburg2025technical,viththarachchige2026concerns} \\
Large-load reliability and regulation & Emerging large-load risks, modeling gaps, interconnection, commissioning, and federal proceedings. & Integrates official reliability and regulatory developments with peer-reviewed technical mechanisms and jurisdictional limits. & \cite{force2025characteristics,ref244,ref248,ref259,ref267} \\
\bottomrule
\end{tabularx}
\end{table}

\section{Data center operations and load characteristics}
\label{sec:operations}

\subsection{Load mix and facility coupling}

Facility demand comprises information technology (IT) equipment, cooling, power-conversion stages, UPS systems, backup-generation interfaces, lighting, controls, and auxiliary services. Their shares vary with server utilization, cooling technology, rack density, climate, and electrical architecture \cite{ahmed2021review,dayarathna2015data,ref14}. A universal load-mix pie chart is therefore misleading. AI-oriented designs tend to increase rack density, accelerator concentration, and the coupling between IT loading and cooling demand, but the magnitude remains facility-specific \cite{chen2025electricity,huntington2025800}.

At the facility boundary, a compact accounting identity is
\begin{equation}
P_{\mathrm{DC}}(t)=P_{\mathrm{IT}}(t)+P_{\mathrm{cool}}(t)+P_{\mathrm{loss}}(t)+P_{\mathrm{aux}}(t),
\label{eq:facility-balance}
\end{equation}
where $P_{\mathrm{DC}}$ is facility demand before explicit dispatch of on-site generation or storage, $P_{\mathrm{IT}}$ is productive computing power, $P_{\mathrm{cool}}$ is thermal-management demand, $P_{\mathrm{loss}}$ includes transformer, UPS, power-distribution-unit, and conversion losses, and $P_{\mathrm{aux}}$ includes controls, security, fire protection, lighting, and other balance-of-plant demand \cite{dayarathna2015data,9984210,gyang2025dynamic}. Eq.~\eqref{eq:facility-balance} is an accounting identity, not a dynamic load model. Grid exchange requires separate signed terms for generation, charging, and discharging.

Component shares are case-dependent. A dynamic model of a 10~MW facility reports a prior estimate of approximately 60\% IT energy use and 30\% cooling energy use. At the study's selected peak operating point, its simulated server-room and heating, ventilation, and air-conditioning components account for about 85\% and 15\% of facility power, respectively \cite{gyang2025dynamic}. The contrast shows why fixed pie charts can misstate both energy use and grid-facing dynamics. Illustrative next-generation AI electrical-architecture roadmaps likewise describe technology-specific scenarios rather than fleet-wide averages \cite{huntington2025800}.

Power usage effectiveness remains a useful energy-efficiency metric, but it is not a complete grid-impact metric. It is widely defined as the ratio of total facility power to IT equipment power \cite{dayarathna2015data,ref14},
\begin{equation}
\mathrm{PUE} = \frac{P_{\mathrm{facility}}}{P_{\mathrm{IT}}},
\label{eq:pue}
\end{equation}
where $P_{\mathrm{facility}}$ is total facility power and $P_{\mathrm{IT}}$ is IT equipment power. A lower PUE indicates less non-IT overhead, but says nothing directly about ramp rate, voltage sensitivity, harmonic spectrum, disturbance ride-through, load rejection, or reconnection. Facilities with similar annual electricity use may therefore impose different grid requirements because their converter topology, UPS mode, workload scheduling, cooling control, and interconnection strength differ \cite{ref99,9984210,chen2025electricity}.

Internal efficiency and grid compatibility are distinct attributes. A low-PUE facility may still require substantial network reinforcement or dynamic study if it connects at a weak location, changes power rapidly, or has uncertain disturbance response. Conversely, a less efficient site may present a more predictable interface. Utility assessment should pair energy metrics with maximum and minimum demand, coincidence, ramping, active and reactive response, power quality, ride-through, reconnection, and controllable flexibility \cite{force2025characteristics,ref244,ref248}.

\subsection{Workload stages and electrical variability}

Computational demand is shaped by workload stage. Model preparation, training, fine-tuning, and inference differ in duration, accelerator utilization, synchronization, storage activity, and tolerance for delay. Training can sustain high utilization yet exhibit periodic communication or checkpointing transients; inference follows user demand more closely and can be bursty; data preparation and fine-tuning often occupy intermediate regimes \cite{chen2025electricity,kwon2025operational,choukse2025power}. These patterns are not universal because model architecture, cluster size, scheduling, and hardware change the electrical profile.

Hardware and software controls mediate the conversion from workload to electrical demand. Power capping, accelerator scheduling, workload shaping, cooling control, and rack-level storage can modify ramps and peaks \cite{chen2025electricity,choukse2025power,tao2025fast}. Such controls create useful flexibility, but also make the load partly dependent on internal policies that may be invisible to the utility unless models and telemetry are exchanged.

For power-system analysis, flexibility is best separated into deferrable energy, fast power modulation, and interruptible service. Deferrable energy moves nonurgent computation over hours or days and is bounded by deadlines and workload placement \cite{ref16,takci2025data,chen2026defer}. Fast modulation changes power over seconds or minutes without necessarily reducing total energy; storage-compute coordination and tie-line smoothing studies illustrate this distinct service \cite{liu2026watts,tao2025fast,ref219}. Interruptible or flexible-interconnection service permits curtailment under specified conditions and therefore requires explicit baselines, notice, recovery, and nonperformance terms \cite{kahrl2026speed,ref259}.

\subsection{Power electronics, UPS systems, cooling, and internal architecture}

At the utility interface, a data center appears through transformers, switchgear, rectifiers, power-factor-correction circuits, UPS systems, power distribution units, server power supplies, cooling drives, and sometimes on-site storage or generation. This chain differs from conventional motor-dominated demand because active and reactive response depend on converter controls, current limits, protection, and operating mode \cite{9984210,sun2022dynamic,li2025ai}. Depending on those controls, aggregate demand can exhibit near-constant-power behavior over part of the voltage range and then change sharply when current limits, transfer logic, or protection thresholds are reached \cite{sun2022dynamic,li2025ai}.

UPS systems preserve critical computing service during supply disturbances and may also provide grid support when controls, battery state, and service obligations allow. Simulation, controller-hardware-in-the-loop, and demonstration studies report potential for frequency response \cite{alapera2018data,tao2025fast}, dynamic active-power support and smoothing \cite{ross2026using}, line-interactive operation \cite{azizi2026strengthening}, and flexible-grid-asset operation on a national-laboratory test platform \cite{vaidhynathan2025vulcan}. These results establish technical potential rather than universal availability; backup reserve, duration, cycling, and customer reliability remain binding constraints.

Cooling usually changes more slowly than compute demand because of thermal inertia, while pumps, fans, chillers, and variable-speed drives add motor and converter dynamics. Thermal storage and supervisory control can decouple part of the cooling demand from short IT-load variations \cite{ref14,winick2025reducing}, but cooling equipment may also add harmonics and weather-sensitive peaks \cite{9176824,mousavi2017power}. Cooling dynamics belong in the load model when the study concerns feeder loading, electromechanical response, or power quality \cite{ahmed2021review,gyang2025dynamic}. Table~\ref{tab:load-characteristics} relates facility characteristics to grid mechanisms.

\begin{table}[!t]
\centering
\tabfont
\setlength{\tabcolsep}{4pt}
\renewcommand{\arraystretch}{1.1}
\caption{Data center load characteristics and grid-impact mechanisms. The time scales indicate dominant planning horizons or study resolutions and are not exclusive bounds.}
\label{tab:load-characteristics}
\begin{tabularx}{\textwidth}{>{\raggedright\arraybackslash}p{0.16\textwidth} Y Y >{\raggedright\arraybackslash}p{0.13\textwidth}}
\toprule
\textbf{Characteristic} & \textbf{Grid-relevant mechanism} & \textbf{Affected layer and time scale} & \textbf{Key sources} \\
\midrule
High power density & Large point load can exceed local transformer, feeder, substation, or transmission hosting capability. & Transmission, subtransmission, distribution; years to months. & \cite{ref3,quint2025practical,krishnamoorthy2026role,chen2025electricity,ref253} \\
Power-electronic interface & Nonlinear current, fast control response, harmonic emission, resonance interactions, and converter-dependent disturbance response. & Point of interconnection, feeder, protection; sub-cycle to seconds. & \cite{9803460,sun2022data,sun2019modeling,li2025ai} \\
Workload-induced variability & Training, inference, synchronization, queueing, and power capping can create ramps or pulses not captured by annual energy forecasts. & Operations, reserves, dynamics; days to seconds. & \cite{chen2025electricity,kwon2025operational,choukse2025power,wang2026regime,tao2025fast} \\
UPS and storage & Ride-through and power smoothing are possible, but transfer, trip, and reconnection behavior must be coordinated. & Point of interconnection and system dynamics; seconds to minutes. & \cite{alapera2018data,xie2025enhancing,alkhudaydi2026safe,tao2025fast} \\
Cooling coupling & Thermal load follows computing demand and may add motor dynamics, harmonic-producing drives, and weather-sensitive peak demand. & Facility, feeder, power quality; minutes to hours. & \cite{ref14,gnibga2026improving,winick2025reducing,9176824,mousavi2017power} \\
Geographic clustering & Large correlated loads can concentrate resource adequacy, transmission, water, and distribution constraints. & Regional planning; years to months. & \cite{ref3,chen2025electricity,ref253,ref267} \\
\bottomrule
\end{tabularx}
\end{table}

\section{Grid impacts across layers and time scales}
\label{sec:grid-impacts}

\begin{figure}[!t]
\centering
\includegraphics[width=\textwidth,height=0.7\textheight,keepaspectratio]{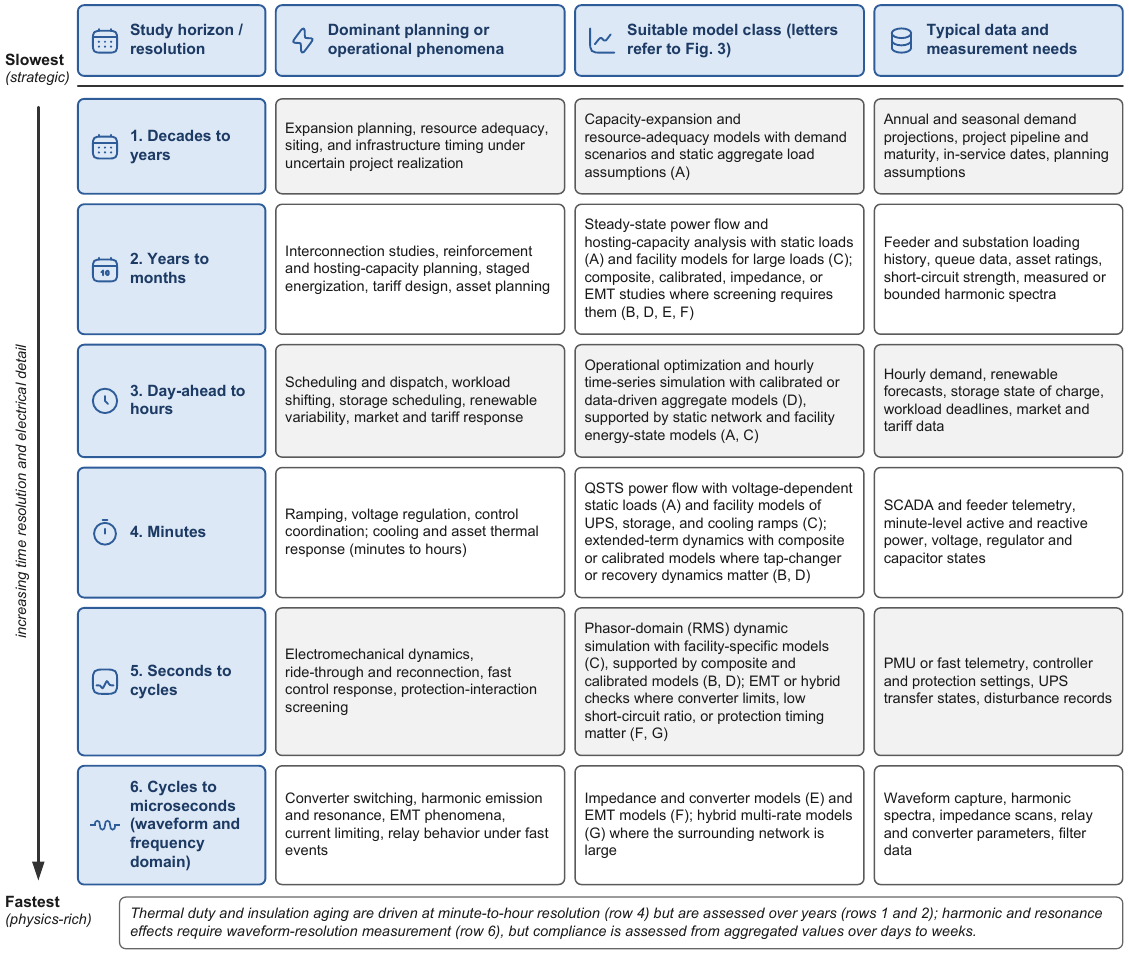}
\caption{Alignment of planning horizons, operating phenomena, and model and data requirements for data center grid studies. The rows denote a decision horizon or required study resolution rather than the duration of every physical effect. Harmonic and resonance phenomena can persist, but their assessment requires waveform or frequency-domain information \cite{chen2025electricity,kwon2025operational,osti_3013288}.}
\label{fig:timescale-taxonomy}
\end{figure}

\subsection{Transmission planning and large-load interconnection}

At bulk-system scale, difficulty arises from the size and pace of new demand. Large-load records and utility assessments describe individual requests ranging from tens to hundreds of megawatts, with some campus proposals larger still \cite{council2024assessment,force2025characteristics,ref253}. Serving such projects may require staged energization, transmission upgrades, generation procurement, and reliability studies \cite{quint2025practical,enriquez2025bridging,kahrl2026speed}. When several requests cluster geographically, uncertain project maturity and overlapping in-service dates can distort forecasts and complicate sequencing \cite{quint2025practical,kahrl2026speed,ref253}.

A credible interconnection study needs more than the requested peak megawatts. Emerging large-load guidance calls for ramp rates, voltage sensitivity, ride-through, power quality, disturbance response, internal segmentation, backup generation, and the interaction between facility protection and the external network \cite{force2025characteristics,ref246,ref248}. Equipment-level studies reach the same conclusion from a different direction: UPS topology and converter controls \cite{zhabelova2015data}, input impedance \cite{sun2019modeling}, and facility-level dynamic behavior \cite{lu2026dynamic,talukdar2026modeling} shape grid response.

Cost allocation and tariff design determine who bears the consequences of a large-load connection. Network and resource upgrades may be assigned to the customer, the utility, or broader ratepayers depending on jurisdiction and service class. Policy analyses emphasize project-maturity tests, cost causation, and protection against speculative or unrealized requests \cite{farmer2025doe,anderson2025policy,kahrl2026speed}. FERC proceedings add federal questions concerning nondiscriminatory transmission access, co-location, forecasting, tariff clarity, and market treatment \cite{ref259,ref264,ref266}.

\subsection{Distribution-system and point of interconnection impacts}

The distribution system is a growing, not a marginal, part of the large-load problem. Many hyperscale campuses connect at transmission voltage through dedicated substations, and a common assumption is that distribution utilities are therefore bystanders. The opposite is increasingly the case: because transmission headroom is limited and existing substations can often be expanded, a substantial share of near-term growth is expected to receive service from shared, mixed-use distribution systems, and distribution utilities are seeing a marked increase in interconnection requests from data centers and other large loads. A 2026 national assessment estimates that roughly 9--10~GW of the current United States project pipeline could feasibly interconnect at distribution voltages, that requests now range from edge facilities above 2~MW to campuses exceeding 100~MW, and that meeting this demand could require on the order of 3000--4000 new feeders and several thousand new substation transformers in the 13--34.5~kV classes \cite{krishnamoorthy2026role}. Large campuses may therefore connect through transmission, subtransmission, or dedicated distribution-like facilities at a bulk-system interface, while smaller and midsize sites, campus distribution networks, and downstream auxiliaries remain within utility distribution scope \cite{krishnamoorthy2026role,force2025characteristics}. Distribution studies show that data center demand and flexibility may alter feeder reliability \cite{fan2025reliability}, phase balance \cite{wang2017phase}, voltage profiles \cite{yan2026switching}, and post-fault reconfiguration and hosting capacity \cite{qiu2026characterizing}.

Distribution effects require a different evidence base from transmission planning. Assets are more numerous, observability is often lower, and equipment is electrically closer to other customers. Direct hyperscale measurements at distribution points of interconnection remain scarce. Transformer studies establish harmonic-heating and derating mechanisms \cite{mitra2023analyzing,bishop1996evaluating,faiz2015derating}; cable studies establish frequency-dependent ampacity and thermal constraints \cite{demoulias2007ampacity,elsharkawy2022effect,rasoulpoor2017thermal}; and computer-load studies establish triplen-harmonic and aggregation effects in neutral conductors \cite{gruzs1990survey,desmet2003analysis,mansoor1995predicting}. These sources establish physical pathways, not universal magnitudes for every facility. Static planning models may also miss converter controls and disturbance behavior \cite{viththarachchige2026concerns,reyes2024effects}.

Distributed energy resources (DERs) change the feeder problem from one-way loading to time-varying net exchange. High photovoltaic penetration can cause voltage rise, variability, and reverse power flow that require coordinated voltage control \cite{padullaparti2022evaluation}. Where export is permitted, a campus with photovoltaic or wind generation and storage may import heavily during compute peaks and export during maintenance, low utilization, or high renewable output \cite{ref14,amin2023development}. Import and export envelopes depend on coincidence, storage state, export limits, feeder impedance, and utility and behind-the-meter controls. Co-location therefore still requires analysis of reverse flow, regulator duty, contingencies, and, where relevant, harmonic and protection conditions.

Hierarchical voltage control, active-network management, and DERMS or ADMS coordination can increase usable hosting capacity or enforce export limits, although results remain feeder- and scenario-specific \cite{castaneda2022electric,strezoski2023distributed,padullaparti2022evaluation}. Data center demand may absorb local renewable production in some hours and worsen peak loading in others. Joint planning should consider facility demand, renewable output, storage, voltage control, and network reinforcement together rather than evaluate load and supply portfolios independently \cite{amin2023development,ref14,wang2017phase}. Effective operation also depends on observability, communications, model fidelity, and enforceable operating rules \cite{wang2015guidelines,ref245}.

\subsection{Markets, tariffs, and operational flexibility}

Data center demand affects wholesale and retail markets through energy use, coincident peaks, capacity obligations, congestion, and reserve requirements. Workload scheduling and geographic shifting can reduce cost or congestion under specified assumptions \cite{ref16,takci2025data,chen2026defer}. Storage-compute coordination can reduce peaks or tie-line variability \cite{alapera2018data,liu2026watts,ref219}. Emissions outcomes remain conditional on the marginal generation mix and the time and destination of shifted demand \cite{evans2025sustainable,fattahi2020energy}. Service-level agreements, latency, workload value, cooling, and recovery constrain the available response.

Market design should ask which service a facility can provide, with what notice, duration, verification, and opportunity cost. The relevant products include load shifting, power capping, storage response, interruptible service, and flexible interconnection, not a binary classification of the site as flexible or inflexible. Technical studies \cite{chen2026defer,mughees2026ai} and policy records \cite{kahrl2026speed,ref259,ref266} increasingly pair these products with improved forecasting, explicit tariff provisions, and performance obligations.

Aggregation creates both value and common-mode risk. A common operator can coordinate geographically dispersed facilities, but shared scheduling, firmware, or protection logic may also synchronize recovery, disconnection, or reconnection. These outcomes are plausible coordination risks rather than universally observed behavior. Programs should specify telemetry, recovery ramps, rebound management, and performance verification \cite{takci2025data,al2025programmable,ref219}.

\subsection{Real-time dynamics, ride-through, and stability}

On sub-second to minute time scales, data centers interact with voltage and frequency through converter controls, UPS transfers, protection, load tripping, and workload modulation. Facility-level dynamic models represent aggregate UPS, cooling, converter, and workload behavior \cite{sun2022dynamic,gyang2025dynamic,lu2026dynamic}. Impedance and resonance studies address frequency-dependent converter-network interactions \cite{sun2022data,zhao2026dynamic,fan2026replicating}. EMT and ride-through models are needed when switching, protection, current limiting, or reconnection transients can change the conclusion \cite{osti_3013288,xie2025enhancing,talukdar2026modeling}.

System risk depends on how the facility behaves during and after a disturbance: whether it stays connected, how active and reactive power change, whether internal blocks transfer or trip together, and how quickly demand returns. NERC guidance and alerts identify modeling, instrumentation, commissioning, protection, and control as immediate areas for computational loads \cite{ref244,ref246,ref248}. Technical studies address the same issues through ride-through, reconnection, and stability analysis \cite{xie2025enhancing,alkhudaydi2026safe,conto2025texas}.

Ride-through is not a binary attribute. Relevant parameters include voltage and frequency thresholds, delay, active and reactive response, recovery shape, internal segmentation, backup generation, UPS energy state, and reconnection ramp. Studies of low-voltage ride-through \cite{xie2025enhancing}, safe reconnection \cite{alkhudaydi2026safe}, large power electronic loads \cite{conto2025texas}, and fast frequency response \cite{tao2025fast} show that facility controls and grid requirements must be designed together. Table~\ref{tab:grid-impacts} summarizes the principal impact categories by grid layer and time scale.

\begin{table}[!t]
\centering
\tabfont
\setlength{\tabcolsep}{4pt}
\renewcommand{\arraystretch}{1.1}
\caption{Grid-impact categories by layer and time scale}
\label{tab:grid-impacts}
\begin{tabularx}{\textwidth}{>{\raggedright\arraybackslash}p{0.16\textwidth} Y Y >{\raggedright\arraybackslash}p{0.13\textwidth}}
\toprule
\textbf{Layer or function} & \textbf{Main impact} & \textbf{Relevant study data} & \textbf{Key sources} \\
\midrule
Bulk system planning & Resource adequacy, transmission expansion, regional clustering, capacity and energy forecasts. & Load growth scenarios, in-service dates, location, coincident peak, expected flexibility. & \cite{ref3,ref9,council2024assessment,ref253} \\
Large-load interconnection & Study queues, upgrade scope, staged energization, cost allocation, ride-through and telemetry requirements. & Peak MW, ramp rate, voltage sensitivity, power factor, protection settings, commissioning tests. & \cite{quint2025practical,force2025characteristics,ref246,ref248,ref259} \\
Distribution feeder & Voltage regulation, phase balance, hosting capacity, local reliability, DER coordination. & Feeder topology, transformer loading, regulators, capacitor banks, local DER, load spectra. & \cite{fan2025reliability,wang2017phase,krishnamoorthy2026role,qiu2026characterizing} \\
Point of interconnection & Power quality, ride-through, reconnection, active and reactive response. & Voltage and current waveforms, phasors, harmonics, event logs, UPS state, telemetry. & \cite{9803460,xie2025enhancing,alkhudaydi2026safe,ref248} \\
Behind-the-meter campus & Workload scheduling, UPS control, storage dispatch, cooling, backup generation. & Facility operating state, workload class, battery energy storage system (BESS) state of charge, cooling mode, internal segmentation. & \cite{ref16,alapera2018data,takci2025data,liu2026watts,tao2025fast} \\
\bottomrule
\end{tabularx}
\end{table}

\section{Power quality and distribution asset degradation}
\label{sec:asset-degradation}

\subsection{Harmonics, resonance, and interpretation of power quality metrics}

Nonlinear conversion occurs at several voltage levels within a data center, including server power supplies, power-factor-correction rectifiers, UPS converters, variable-speed cooling drives, and auxiliary electronics. Measurements at facility distribution points document current and voltage distortion \cite{9692449,9803460}, while cooling-drive and sensor studies show that emission and observability depend on subsystem and measurement location \cite{9176824,portillo2026monitoring}. Impedance studies show that facility and network impedances can amplify selected frequencies \cite{sun2022data,sun2019modeling,fan2026replicating}. Risk therefore depends jointly on the emitted spectrum and the frequency-dependent network seen from the point of common coupling (PCC).

A standard descriptor used in the reviewed measurement literature is current total harmonic distortion (THD),
\begin{equation}
\mathrm{THD}_{I}=\frac{\sqrt{\sum_{h=2}^{H} I_h^2}}{I_1}\times 100\%,
\label{eq:thd}
\end{equation}
where $I_h$ is the root-mean-square current at harmonic order $h$, $I_1$ is the fundamental component, and $H$ is the highest order retained \cite{9803460,portillo2026monitoring,nassif2018power}. Eq.~\eqref{eq:thd} reports aggregate distortion but is not, by itself, a compliance, asset-health, or resonance metric. IEEE 519-2022 establishes steady-state distortion limits at the point of common coupling: voltage-distortion limits, which are the responsibility of the system owner, and current-distortion limits, which are the responsibility of the user \cite{ieee519_2022}. Because the denominator in Eq.~\eqref{eq:thd} is the contemporaneous fundamental current, comparisons across operating points and compliance assessments must use the metric, normalization, point-of-common-coupling definition, and individual harmonic limits required by the applicable standard. Consequence also depends on harmonic order and sequence, phase-angle diversity, loading, conductor geometry, transformer construction, thermal boundary conditions, and network impedance.

These findings show why a scalar distortion index is insufficient. Triplen harmonics from single-phase nonlinear loads may accumulate in the neutral, although device diversity, attenuation, and phase-angle differences prevent a universal arithmetic sum \cite{gruzs1990survey,desmet2003analysis,mansoor1995predicting}. Harmonic currents increase frequency-dependent conductor, winding-eddy, and stray losses in cables and transformers \cite{bishop1996evaluating,faiz2015derating,tofoli2006study}. Because THD collapses the spectrum into one value, it cannot by itself rule out a narrow-band interaction between facility converters and upstream impedance \cite{sun2022data,ruan2025data,fan2026replicating}. Screening should therefore combine harmonic spectra with frequency-dependent impedance or resonance analysis.

\subsection{Transformer heating, derating, and insulation aging}

The transformer literature combines studies of increased power-electronic penetration with broader nonlinear-load research. The former identify additional losses and thermal duty and motivate condition-informed management \cite{mitra2023analyzing,nair2026predictive}. The broader literature shows that sustained loading raises top-oil and winding temperatures, while harmonic current increases winding-eddy and other stray losses \cite{bishop1996evaluating,faiz2015derating}; the resulting hot-spot temperature affects insulation-aging rate \cite{said2010analysis,cazacu2018thermal}. Most thermal and derating studies are not field measurements from hyperscale service transformers. They establish transferable mechanisms whose magnitude must be determined from facility-specific spectra, loading, construction, and cooling.

For a winding resistance $R_{\mathrm{dc}}$ evaluated at the relevant winding temperature, the ohmic component of winding loss under a nonsinusoidal current is
\begin{equation}
P_{\mathrm{ohm}}=R_{\mathrm{dc}}\sum_{h=1}^{H} I_h^2,
\label{eq:harmloss}
\end{equation}
where the harmonic currents $I_h$ are expressed as root-mean-square components. Eq.~\eqref{eq:harmloss} represents only the direct-current-resistance component of winding loss. A transformer assessment must additionally model no-load or core loss, winding-eddy loss, other stray loss, harmonic-order weighting, ambient and cooling conditions, thermal time constants, and the applicable insulation-aging relationship \cite{faiz2015derating,tofoli2006study,cazacu2018thermal}. Harmonic-loss factors and K-factor-based specifications can support the evaluation of applicable transformer designs, but they do not replace manufacturer data or design-specific thermal analysis.

Peak megawatts and total root-mean-square current are inadequate when spectral content is material. The reviewed methods support a sequence of spectrum measurement or estimation, harmonic-loss calculation, thermal modeling, and loading-margin or remaining-life assessment \cite{bishop1996evaluating,said2010analysis,faiz2015derating}. Where spectra are uncertain, analyses should use bounded sensitivity cases rather than assume either worst-case addition or cancellation. Transformer selection and loading limits should reflect the expected spectrum, cooling arrangement, contingency duty, and planned equipment changes.

\subsection{Cable ampacity, underground thermal constraints, and neutral conductors}

Low-voltage cable studies show that nonsinusoidal current can increase phase-conductor and neutral losses, with sensitivity to conductor size, installation method, phase allocation, and zero-sequence content \cite{demoulias2007ampacity,hiranandani1998calculation,meliopoulos1992calculation}. Underground medium-voltage studies add sheath loss, burial arrangement, soil thermal resistivity, mutual heating, and load-cycle effects \cite{elsharkawy2022effect,rasoulpoor2017thermal}. These effects are route- and design-specific; they do not imply one universal harmonic derating factor.

Neutral conductors need separate assessment where many single-phase computer-type supplies are present. Balanced fundamental phase currents do not guarantee low neutral current because triplen harmonics are zero-sequence quantities. Surveys and aggregation studies report both high neutral currents and substantial sensitivity to device diversity, phase angles, and attenuation \cite{gruzs1990survey,desmet2003analysis,mansoor1995predicting}. Design should use measured or equipment-derived spectra and actual phase allocation rather than assume complete addition or cancellation.

Route-specific thermal studies show that underground circuits can become a limiting element under long duct banks, adverse soil conditions, multiple loaded circuits, or significant sheath loss \cite{elsharkawy2022effect,rasoulpoor2017thermal}, and that zero-sequence harmonic current further erodes the margin \cite{hiranandani1998calculation,meliopoulos1992calculation}. Applying that result to a data center campus is a planning inference rather than a universal field observation. Ampacity studies should therefore include the complete route, neutral or sheath paths, installation environment, load cycle, and expected harmonic duty.

\subsection{Protection, metering, and monitoring implications}

Direct field documentation of data center-induced protection misoperation remains limited. Rectifier and adjustable-speed-drive studies document distorted current and associated equipment or measurement effects \cite{rice1986adjustable}. Reviews of converter-dominated protection identify control-dependent fault-current magnitude, sequence content, and duration \cite{cao2023review,nagpal2020protection}. These are engineering analogies, not proof that a given data center will produce the same relay response. NERC's computational-load alert is more direct in calling for study, instrumentation, commissioning, protection, and control actions for large computational loads \cite{ref248}.

Risk-based monitoring can narrow this uncertainty. Depending on facility size, voltage level, and network strength, useful records include synchronized active and reactive power, voltage and current waveforms, harmonic spectra, event and protection logs, UPS state, transformer thermal estimates, feeder topology, and commissioning results \cite{chatterjee2026measurement,portillo2026monitoring,ref248}. The required package is study-specific; not every installation needs every measurement at the same resolution.

The principal gap is the integration of established physical methods with data center field measurements. Harmonic heating, ampacity, and neutral-current models are mature in adjacent nonlinear-load literature \cite{bishop1996evaluating,demoulias2007ampacity,gruzs1990survey}, whereas current large-load guidance concentrates on capacity, modeling, ride-through, commissioning, and operational coordination \cite{force2025characteristics,ref246,ref248}. Few studies link measured facility spectra to feeder-asset aging and utility investment decisions. Table~\ref{tab:asset-degradation} distinguishes established mechanisms from their present applicability to data center planning.

\subsection{Harmonic-constrained optimal power flow for converter-dominated loads}

Conventional optimal power flow (OPF) formulations regulate fundamental-frequency quantities such as active and reactive power, voltage magnitude, and thermal loading, without representing the waveform distortion produced by nonlinear and converter-interfaced equipment. In networks with substantial harmonic injection, a dispatch that is feasible at the fundamental frequency may still violate voltage or current distortion limits, increase transformer and cable losses, disturb shunt capacitor banks or excite filter resonances, and shorten equipment life. Harmonic-constrained optimal power flow (H-OPF) therefore co-optimizes operating set points while enforcing harmonic power quality requirements such as those of IEEE 519-2022 \cite{ieee519_2022}, using harmonic power-flow representations of the kind described in IEEE 3002.8-2018 \cite{ieee3002_8_2018}. Formulations that embed IEEE 519 distortion limits in planning and dispatch problems have been developed mainly for inverter-based distributed generation, where they determine the permissible penetration, location, and size of converter-interfaced sources under harmonic and protection constraints \cite{pandi2013determining,pandi2013optimal}.

This framing is directly relevant to large data centers, and increasingly to AI-oriented facilities. Although accelerators operate internally from regulated direct-current supplies, their aggregate demand is served through extensive power-electronic infrastructure: high-capacity UPS rectifier and inverter stages, active power-factor-correction (PFC) converters in server power supplies, direct-current conversion stages supplying accelerator racks, battery-storage interfaces, and variable-frequency drives (VFDs) for chillers, pumps, cooling towers, and air-handling equipment \cite{9984210,li2025ai}. The simultaneous operation and rapid loading variation of these devices can alter harmonic injections and create frequency-dependent interactions with transformers, cables, capacitor banks, and harmonic filters \cite{sun2022data,fan2026replicating}. The concern is amplified by the rapid expansion of data-center electricity demand associated with AI workloads \cite{ref3,shehabi2024report}.

Significant gaps remain for data-center applications. Existing harmonic-constrained formulations generally assume balanced networks, fixed harmonic current spectra, and static load conditions, whereas AI data centers exhibit phase imbalance, fast workload-driven demand changes, uncertain equipment aggregation, and proprietary converter controls. Further research is needed on scalable, measurement-informed, unbalanced, and multi-time-scale H-OPF frameworks that jointly represent fundamental-frequency operation, harmonic and resonance constraints, uncertainty in converter emissions, and the distinction between conventional harmonics and higher-frequency supraharmonic emissions \cite{ieee519_2022,ieee3002_8_2018,bollen2006signal}.

\begin{table}[!t]
\centering
\tabfont
\setlength{\tabcolsep}{4pt}
\renewcommand{\arraystretch}{1.1}
\caption{Distribution asset-stress mechanisms, applicability, and planning implications for data center loads}
\label{tab:asset-degradation}
\begin{tabularx}{\textwidth}{>{\raggedright\arraybackslash}p{0.15\textwidth} Y Y Y}
\toprule
\textbf{Asset or function} & \textbf{Established mechanism} & \textbf{Applicability and limitation} & \textbf{Planning implication and key sources} \\
\midrule
Transformer & Harmonic winding-eddy and stray losses combine with sustained loading to raise hot-spot temperature and affect insulation aging. & Mainly nonlinear-load and power-electronic-load studies; limited hyperscale field aging data. & Use measured or bounded spectra, design-specific loss allocation, and thermal modeling rather than MW alone \cite{mitra2023analyzing,nair2026predictive,bishop1996evaluating,faiz2015derating,cazacu2018thermal}. \\
Low-voltage cable and neutral & Frequency-dependent conductor loss and zero-sequence harmonic current reduce thermal margin. & Computer-load and generic nonlinear-load studies provide transferable mechanisms. & Check phase allocation, neutral sizing, installation, and harmonic ampacity \cite{gruzs1990survey,demoulias2007ampacity,desmet2003analysis,mansoor1995predicting}. \\
Medium-voltage underground cable & Conductor and sheath losses interact with daily cycling, soil, burial, and circuit geometry. & Cable-thermal studies, not direct data center campus field campaigns. & Include duct-bank and soil conditions in interconnection ampacity studies \cite{patil2011effects,elsharkawy2022effect,rasoulpoor2017thermal}. \\
Protection and metering & Distorted waveforms and control-dependent converter response can affect measurement and protection assumptions. & Mostly converter-dominated resource analogies; direct data center relay evidence is sparse. & Require model review, commissioning tests, disturbance recording, and case-specific protection studies \cite{cao2023review,nagpal2020protection,ref248}. \\
Asset management & Monitoring links load spectra and thermal duty to remaining-life estimates and upgrade timing. & Predictive methods exist, but utility deployment for large computational loads is not well documented. & Combine power quality records, loading history, thermal estimates, and model updates \cite{nair2026predictive,portillo2026monitoring,culler2026digital,said2010analysis}. \\
\bottomrule
\end{tabularx}
\end{table}

\section{Modeling requirements for data center loads}
\label{sec:modeling}

\begin{figure}[!t]
\centering
\includegraphics[width=\textwidth,height=0.7\textheight,keepaspectratio]{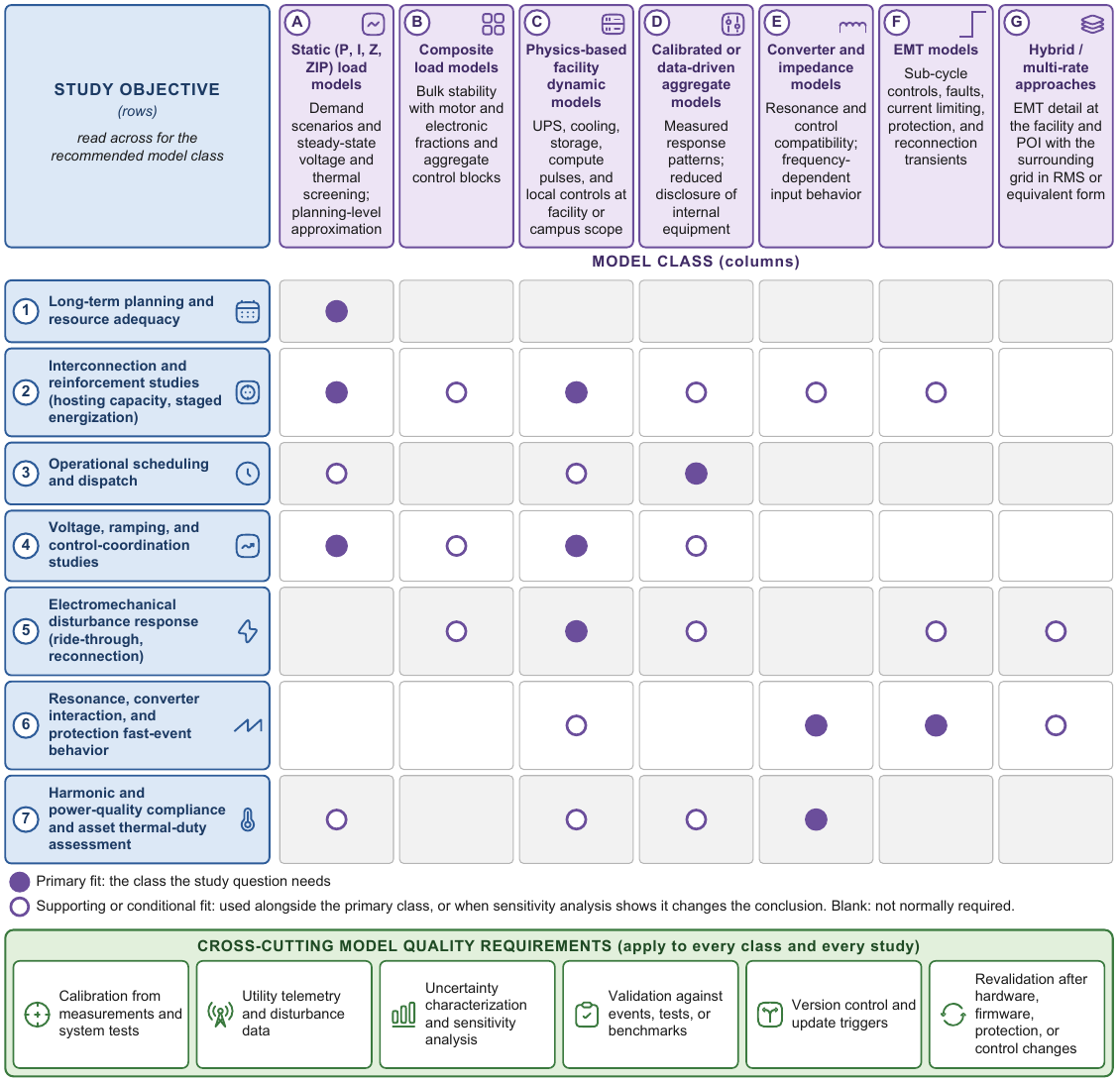}
\caption{Study-oriented model-selection framework for data center loads. The model classes are not a universal ranking; selection follows the study question, time scale, network strength, protection behavior, data availability, and required uncertainty. Measurement, calibration, version governance, and revalidation after hardware or control changes apply across the model portfolio \cite{gyang2025dynamic,osti_3013288,ref248}.}
\label{fig:modeling-hierarchy}
\end{figure}

Figure~\ref{fig:modeling-hierarchy} organizes model classes by the question they can answer. More detail is not automatically better. The appropriate fidelity follows from the time scale, network strength, protection behavior, data availability, and decision being supported.

\subsection{What static models can and cannot represent}

Constant-power, constant-current, and constant-impedance models, and their combination as the constant impedance-current-power (ZIP) model, remain useful for fundamental-frequency power-flow screening and scenario analysis. A conventional static voltage-dependence curve, however, cannot represent discrete UPS states, converter current limits, protection thresholds, reconnection logic, harmonic emission, storage state, or workload pulses \cite{viththarachchige2026concerns,reyes2024effects,ref248}. This is a limitation of the model class, not a claim that every facility exhibits every behavior.

Static models can identify steady-state thermal and voltage constraints, but they cannot determine whether a facility rides through a sag, transfers to storage, trips in blocks, excites a mode, or reconnects with an excessive ramp. Dynamic models of data centers address aggregated operating states \cite{sun2022dynamic,lu2026dynamic}; impedance and resonance models address modal interactions \cite{fan2026replicating}; and EMT or transient-stability models address fast ride-through and reconnection behavior \cite{osti_3013288,xie2025enhancing,alkhudaydi2026safe}. Facilities with the same net demand may need different dynamic representations because UPS topology, protection, converter controls, cooling composition, and internal segmentation differ \cite{ref111,azizi2026strengthening,reyes2024effects}.

\subsection{Dynamic, calibrated, impedance, EMT, and hybrid representations}

Physics-based aggregate models represent selected internal states such as UPS or storage response, cooling motors, electronic load fractions, and pulsed computing demand. They support phasor-domain stability and ride-through studies when parameters are available, but aggregation may hide control interactions and protection boundaries \cite{gyang2025dynamic,osti_3013288,talukdar2026modeling}. Calibrated or pattern-consistent models identify aggregate behavior from measurements and may reduce disclosure of internal equipment. Their validity can deteriorate when hardware, workload, or control settings change \cite{lu2026dynamic}.

Converter and impedance models address compatibility between the input behavior of power supplies or UPS stages and network impedance. They are suited to resonance and control-stability questions but require frequency-dependent parameters and defensible aggregation \cite{sun2022data,sun2019modeling,zhao2026dynamic}. They do not replace a facility-level ride-through model because the two representations answer different questions.

EMT models are warranted when sub-cycle switching, converter controls, current limiting, protection, detailed fault response, or reconnection transients could change the conclusion. They are not necessary for every planning study. Their data and computational burden motivate hybrid approaches that retain detail at the facility and point of interconnection while reducing the surrounding network or less influential internal equipment \cite{osti_3013288,li2025ai,reyes2024effects}. Sensitivity analysis should verify that the reduced portion does not remove the interaction of interest.

\subsection{Calibration, data exchange, and model governance}

Model validity depends on commissioning data and continued maintenance. Workload, equipment, protection, and control information may be commercially or security sensitive, while utilities still need enough behavior to assess stability, power quality, restoration, and operating limits. The literature supports aggregated or privacy-preserving exchange and digital assurance \cite{ref21,culler2026digital}, combined with standard templates, commissioning tests, and disturbance-based validation \cite{chatterjee2026measurement,ref246,ref248}.

The minimum data set follows the study purpose. Resource-adequacy analysis needs credible project maturity, energization dates, and demand scenarios. Power-flow and hosting-capacity studies need location, maximum and minimum demand, reactive capability, and voltage sensitivity. Dynamic studies need active and reactive response, protection and ride-through states, UPS transfer, segmentation, recovery, and reconnection. Power quality and asset studies need spectra, impedance, filters, loading history, and thermal parameters \cite{osti_3013288,talukdar2026modeling,ref248}.

Governance is needed because a computational facility may change after its interconnection study. New accelerators, power supplies, cooling systems, UPS modes, protection, or control settings may invalidate an accepted model. NERC guidance \cite{ref245} and alerts \cite{ref246,ref248} call for validation, updated information, commissioning, instrumentation, and operational coordination for emerging or computational loads. Table~\ref{tab:model-classes} summarizes the intended use and limitations of the principal model classes.

\begin{table}[!t]
\centering
\tabfont
\setlength{\tabcolsep}{4pt}
\renewcommand{\arraystretch}{1.1}
\caption{Study-oriented comparison of data center load model classes}
\label{tab:model-classes}
\begin{tabularx}{\textwidth}{>{\raggedright\arraybackslash}p{0.15\textwidth} Y Y Y}
\toprule
\textbf{Model class} & \textbf{Study question and minimum representation} & \textbf{Validation data} & \textbf{Principal limitation and key sources} \\
\midrule
Static P, I, Z, or ZIP & Steady-state voltage and thermal screening; represent aggregate active/reactive voltage dependence. & Metered steady-state points or accepted planning assumptions. & Cannot represent discrete controls, protection, harmonics, or recovery \cite{ross2026using,viththarachchige2026concerns,reyes2024effects}. \\
Composite dynamic & Bulk stability with motors, electronic fractions, and aggregate control blocks. & Disturbance response or calibrated component fractions. & Generic blocks can conceal data center-specific UPS and protection behavior \cite{sun2022dynamic,mao2026online}. \\
Physics-based data center dynamic & Ride-through, ramping, and recovery with UPS, cooling, compute pulses, and active/reactive states. & Commissioning tests and point of interconnection event records. & Parameter and aggregation uncertainty; requires updates after facility changes \cite{gyang2025dynamic,osti_3013288,talukdar2026modeling}. \\
Calibrated or data-driven aggregate & Reproduce measured response patterns with reduced disclosure of internal equipment. & Representative operating and disturbance data across regimes. & Extrapolation outside the calibration envelope can be unreliable \cite{lu2026dynamic,chatterjee2026measurement}. \\
Converter or impedance & Resonance and control compatibility; represent frequency-dependent input behavior and network impedance. & Frequency-response or component-model validation. & Does not by itself represent all facility operating states or thermal behavior \cite{sun2022data,sun2019modeling,fan2026replicating,zhao2026dynamic}. \\
EMT or hybrid & Sub-cycle controls, faults, current limiting, protection, and detailed reconnection. & Equipment parameters, controller settings, event waveforms, and interface tests. & High data and computational burden; model boundary must be sensitivity-tested \cite{osti_3013288,li2025ai,reyes2024effects}. \\
\bottomrule
\end{tabularx}
\end{table}

\section{Regulatory, reliability, and interconnection developments}
\label{sec:regulation}

At the time of writing, the North American record had progressed from risk identification to alerts, directed standards development, proposed registration criteria, and regional tariff proceedings. These instruments do not have equal legal force. NERC white papers and reliability guidelines are advisory \cite{force2025characteristics}, and alerts request actions or information \cite{ref248}. FERC orders can direct standards development \cite{ferc2026computationalorder} or open region-specific tariff proceedings \cite{ref263}, but draft registry criteria and standards do not become enforceable until the applicable NERC and FERC processes are completed.

NERC's 2025 white paper identified planning, operational, stability, power quality, observability, cybersecurity, and restoration risks associated with emerging large loads, including computational facilities \cite{force2025characteristics}. Its September 2025 Level 2 recommendation addressed interconnection studies, models, commissioning, ramps, and operational coordination; the 2026 aggregated report found uneven readiness and process maturity \cite{ref246,ref247}. The 2026 gap assessment, mitigation guideline, and Level 3 computational-load alert extended the work to modeling, instrumentation, operations, protection, and control \cite{ref244,ref245,ref248}. The Level 3 alert was issued in May 2026, with registered-entity responses due in August 2026. The proposed Computational Load Entity rationale and Project 2026-02 remained developmental \cite{ref249,ref250}. In July 2026, FERC directed NERC under section 215(d)(5) of the Federal Power Act to file, by the end of 2026, one or more new or modified Reliability Standards addressing computational-load integration, and under section 215(f) to submit by the same deadline revisions to its Rules of Procedure, including registry criteria for computational load entities \cite{ferc2026computationalorder}. The directive accelerates the standards pathway but does not itself make the draft criteria or standards enforceable. It also requires NERC to submit, by March 2027, an informational filing with a workplan for the subsequent (Phase II) standards.

FERC and U.S. Department of Energy records address transmission service, forecasting, tariffs, and co-location. A November 2024 technical conference examined large loads co-located with generation \cite{ref264}. FERC's February 2025 action involving PJM Interconnection sought tariff clarity for AI data center co-location, and a September 2025 letter asked regional operators to improve large-load forecasting \cite{ref265,ref266}. The Secretary of Energy's October 2025 Section 403 proposal was docketed as RM26-4 and treated as an advance notice of proposed rulemaking \cite{ref259}. FERC stated its intent to act in April 2026 and in June 2026 issued six region-specific Section 206 show-cause orders directing regional transmission organizations (RTOs) and independent system operators (ISOs) to justify or reform aspects of their large-load tariffs \cite{ref262,ref263}. Currently, RM26-4 and the six regional proceedings have not established one nationwide final large-load interconnection rule. The Southwest Power Pool concurrence illustrates a load-limited approach and concerns about cost protection and electrically proximate generation, while the FERC staff market report provides context rather than binding Commission policy \cite{ref267,ref268}.

The scope of these records is bounded. NERC assessments address the bulk power system, and the 2025 Long-Term Reliability Assessment does not evaluate lower-voltage distribution reliability \cite{ref253}. FERC proceedings concern federal transmission and wholesale-market jurisdiction. Neither source set replaces distribution engineering, state regulation, local tariffs, international grid codes, or asset-specific studies. They are used here as a North American implementation case that links technical evidence with evolving reliability and tariff practice. Table~\ref{tab:regulatory-developments} summarizes the status and scope of the principal instruments.

\begin{table}[!t]
\centering
\tabfont
\setlength{\tabcolsep}{4pt}
\renewcommand{\arraystretch}{1.1}
\caption{Status, scope, and applicability of selected NERC and FERC large-load developments}
\label{tab:regulatory-developments}
\begin{tabularx}{\textwidth}{>{\raggedright\arraybackslash}p{0.17\textwidth} Y Y Y}
\toprule
\textbf{Date and status} & \textbf{Instrument} & \textbf{Principal issue} & \textbf{Status and limitation} \\
\midrule
2025, NERC white paper & Characteristics and risks of emerging large loads \cite{force2025characteristics}. & Multi-domain risk taxonomy for emerging large loads. & Technical risk framing; not an enforceable standard. \\
Sep. 2025 to Mar. 2026, NERC alert and response report & Level 2 recommendation and aggregated industry report \cite{ref246,ref247}. & Model data, studies, commissioning, ramps, and operational coordination. & Requested actions and industry-readiness evidence. \\
Mar. to May 2026, NERC guidance and Level 3 alert & Gap assessment, mitigation guideline, and computational-load alert \cite{ref244,ref245,ref248}. & Modeling, instrumentation, ride-through, protection, power quality, and operations. & Reliability guidance and alert actions; scope centered on bulk-system reliability. \\
Apr. to Jul. 2026, standards and registration pathway & Proposed Computational Load Entity, Project 2026-02, and FERC's RD26-7 directive \cite{ref249,ref250,ferc2026computationalorder}. & Registry criteria and computational-load Reliability Standards. & Development remains underway; FERC directed filings by the end of 2026, but no new computational-load standard or registry criterion is currently enforceable. \\
Nov. 2024 to Feb. 2025, FERC conference and action & Co-location conference and PJM AI data center action \cite{ref264,ref265}. & Transmission service, resource adequacy, cost allocation, market and customer impacts. & Federal co-location record; not a general finding that co-location eliminates grid dependence. \\
Sep. 2025 to Jul. 2026, federal forecasting and tariff actions & Forecasting letter, RM26-4 advance notice, intent-to-act order, and six regional show-cause actions \cite{ref266,ref259,ref262,ref263}. & Project maturity, study process, flexible service, cost transparency, and tariff reform. & Region-specific proceedings remained active at the time of writing; no nationwide final rule had been issued. \\
2026, FERC staff and Commissioner materials & State of the Markets and SPP concurrence \cite{ref267,ref268}. & Market context and a load-limited interconnection example. & Staff or individual-Commissioner evidence, not a Commission-wide final rule. \\
\bottomrule
\end{tabularx}
\end{table}

\section{Integration strategies and implementation requirements}
\label{sec:solutions}

\begin{figure}[!tp]
\centering
\includegraphics[width=\textwidth,height=0.7\textheight,keepaspectratio]{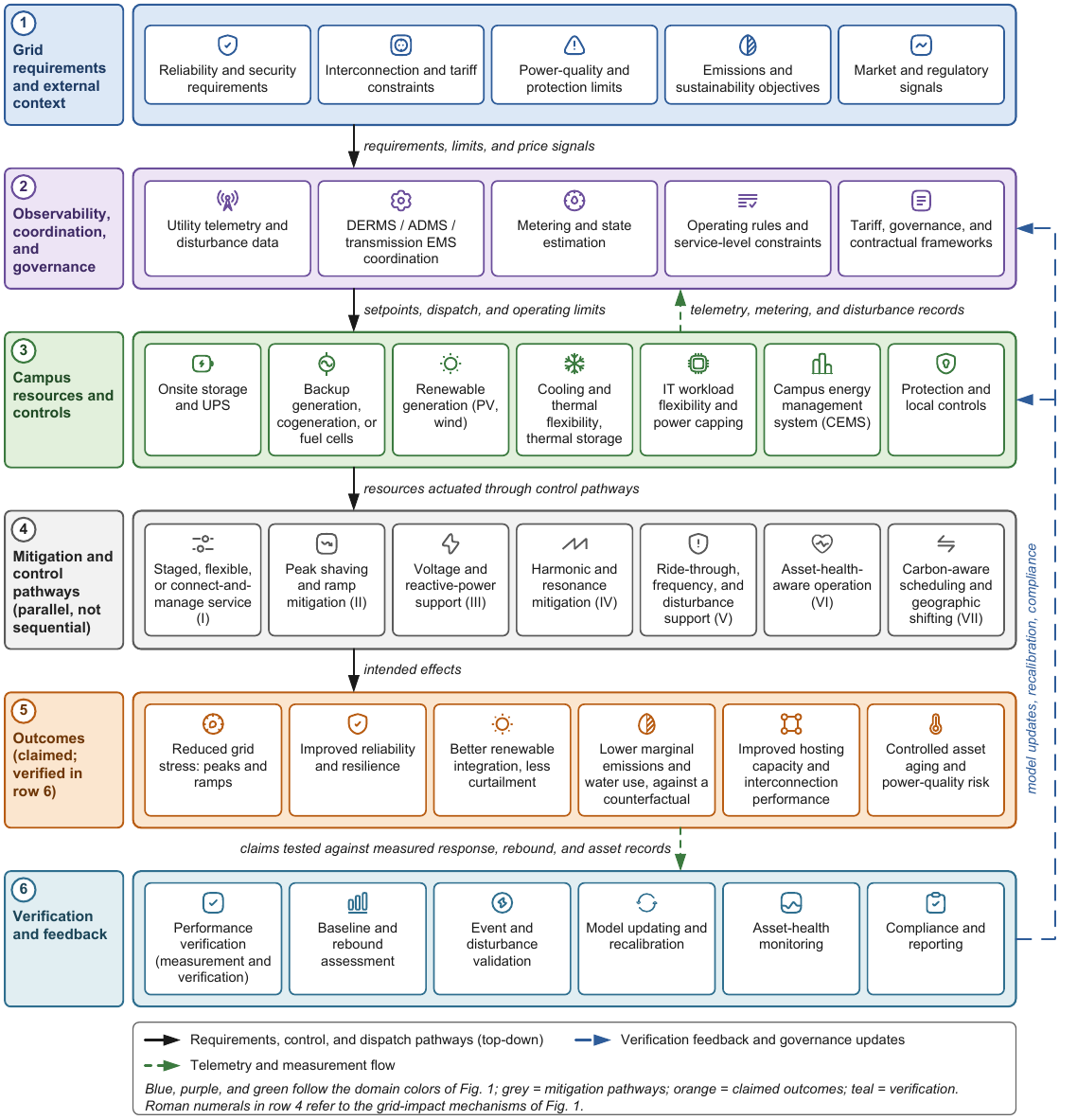}
\caption{Integrated mitigation and verification architecture for grid-compatible data center growth. Grid, market, and reliability requirements (row 1) are connected to campus resources and controls (row 3) through observability, coordination, and governance functions (row 2): utility telemetry and disturbance data, DERMS, ADMS, or EMS coordination, metering and state estimation, operating rules, and tariff and contractual frameworks. Mitigation and control pathways (row 4) lead to the intended outcomes (row 5), which are verified through performance verification, baseline and rebound assessment, event and disturbance validation, model updating and recalibration, asset-health monitoring, and compliance reporting (row 6). Claimed reliability, flexibility, and sustainability benefits should be verified against measured response availability, rebound, asset-health records, model updates, and carbon, water, and lifecycle counterfactuals \cite{strezoski2023distributed,ref155,al2025programmable}.}
\label{fig:solution-architecture}
\end{figure}

Figure~\ref{fig:solution-architecture} groups mitigation options by the failure mechanism they address. Planning, dynamic, power-quality, asset, and sustainability risks require different controls and verification methods. A credible integration strategy combines interconnection practice, validated models, storage or UPS control, workload management, asset monitoring, and supply choices according to network conditions and verified facility capability.

\subsection{Interconnection reform and study data}

A large-load application should separate five information classes. Project-maturity data cover site control, commercial readiness, staged energization, and forecast uncertainty. Steady-state data cover maximum and minimum demand, ramps, reactive capability, and contingency service. Dynamic data cover voltage and frequency sensitivity, ride-through, segmentation, UPS transfer, protection, and reconnection. Power quality data cover spectra, filters, switching states, and relevant impedance. Operational data cover telemetry, communications, commissioning, update obligations, and committed flexibility \cite{force2025characteristics,ref246,ref248}.

Standard templates can reduce omissions and repeated study cycles without imposing identical requirements on every facility. Study scope should follow size, voltage level, network strength, equipment, controllability, and the consequence of common-mode behavior. Firm, flexible, staged, and connect-and-manage service \cite{lu2026battery,chen2026defer,ref155} may accelerate energization only when curtailment rights, telemetry, operating limits, upgrade triggers, and cost allocation are explicit \cite{anderson2025policy,kahrl2026speed}.

\subsection{DERMS, ADMS, and utility coordination}

Distributed energy resource management systems and advanced distribution management systems offer an architecture for coordinating data center demand with feeder limits, voltage control, storage, and local generation. Most supporting studies are generic, simulation-based, or demonstration-oriented rather than evidence of routine production control of hyperscale fleets \cite{strezoski2023distributed,wang2015guidelines,castaneda2022electric}. Their present contribution is therefore architectural and prospective.

Implementation requires an operating agreement that converts planning assumptions into measurable obligations: available reduction or injection, response time, duration, state constraints, communications, measurement and verification, cybersecurity, and fallback behavior. Without this link, a planning study may credit flexibility that the operating platform cannot deliver reliably \cite{takci2025data,chatterjee2026measurement,ref248}. Field demonstrations should test the complete planning-to-operations chain.

\subsection{UPS and storage across time scales}

UPS energy storage is intrinsic to many facilities, while additional battery energy storage systems (BESS), flywheels, supercapacitors, and hybrid systems can extend the range of grid services \cite{ref44,mohammadi2026grid}. Studies report peak reduction and energy shifting \cite{alapera2018data,ref60,lu2026battery}, fast frequency or active-power support \cite{ross2026using,tao2025fast}, and grid-forming or line-interactive operation \cite{azizi2025grid,azizi2026strengthening}.

Storage controls must preserve the facility's primary reliability function. Response capability depends on state of charge, reserved backup energy, power and energy rating, converter limits, duration, thermal constraints, cycling cost, and recharge strategy. Correlated recharge after an event may create a secondary ramp if recovery is not coordinated \cite{ref44,lu2026battery,azizi2026strengthening}. Interconnection studies should state whether storage support is mandatory, contracted, or merely technically possible.

\subsection{Workload flexibility, tariffs, and geographic shifting}

Data center flexibility comprises distinct products: temporal shifting of delay-tolerant work \cite{ref16,takci2025data}, geographic redistribution \cite{chen2026defer,ref219}, and short-term power capping, firm curtailment, and storage-mediated response \cite{liu2026watts}. These products differ in response time, duration, energy recovery, baseline, service-level consequence, and network effect.

Optimization studies demonstrate potential, not universal dispatchability. Latency, service agreements, sunk computing capital, workload value, cooling recovery, data locality, and customer commitments constrain response. Tariffs and contracts should define availability, measurement, compensation, rebound, nonperformance, and recovery \cite{ref155,ref266,ref267}. Geographic shifting also needs destination headroom and network bandwidth; moving computation can transfer congestion or emissions rather than remove them.

\subsection{Co-location, renewable integration, and clean firm supply}

Co-location may combine computational demand with photovoltaic or wind generation, storage, cogeneration, fuel cells, geothermal, nuclear, or other firm supply. For sustainable-energy assessment, physical proximity and annual energy matching are insufficient. Each configuration should be compared with a credible counterfactual using hourly and locational matching, congestion, renewable curtailment, marginal emissions, backup and balancing obligations, network upgrades, asset duty, water and land use, and lifecycle effects \cite{ref14,mohammadi2026grid,ljungblomoptimal}.

On-site photovoltaic or wind generation with BESS may reduce net imports and improve resilience, but performance depends on resource quality, load coincidence, storage duration, export limits, and feeder hosting capacity. A Dublin case study co-optimizes grid imports, photovoltaic generation, wind, storage, and a prospective small modular reactor portfolio; its findings are conditional on the assumed costs, market design, and project scale \cite{ljungblomoptimal}. A Bangladesh photovoltaic-battery microgrid study similarly illustrates a location-specific resilience pathway, not a universal least-cost design \cite{amin2023development}.

Cogeneration or trigeneration can raise total fuel utilization when a useful heat or cooling sink is available. Fuel-cell cogeneration may also provide firm on-site electricity and useful heat \cite{ref212}. Neither option is inherently low-carbon. Net benefit depends on fuel source and upstream emissions, electrical and thermal efficiency, local air quality, water use, capacity factor, heat-sink utilization, and the emissions of displaced grid supply \cite{ref14,siegle2025data}. Natural-gas cogeneration should be reported as a conditional reliability and efficiency pathway, not as renewable supply.

Waste-heat recovery and thermal storage can lower cooling demand or serve an external heat use, but feasibility depends on source temperature, the distance and timing of a heat sink, space, piping, and the energy required for recovery or upgrading. Heat pumps, district-heating links, chilled-water storage, and underground thermal storage illustrate different temperature and duration requirements \cite{ref14,gnibga2026improving,winick2025reducing}. Avoided energy and system emissions are more informative than recovered heat quantity alone.

Co-location also does not establish electrical independence. A facility may continue to use transmission service, ancillary services, backup supply, and system balancing, and may alter deliverability or costs for other customers. The FERC co-location record identifies these issues in the United States, but outcomes depend on network and jurisdiction \cite{ref255,ref264,ref265}. Table~\ref{tab:sustainable-pathways} compares the principal supply and flexibility pathways and the conditions under which their benefits can be claimed.

\begin{table}[!t]
\centering
\tabfont
\setlength{\tabcolsep}{4pt}
\renewcommand{\arraystretch}{1.1}
\caption{Sustainable supply and flexibility pathways for data centers}
\label{tab:sustainable-pathways}
\begin{tabularx}{\textwidth}{>{\raggedright\arraybackslash}p{0.16\textwidth} Y Y >{\raggedright\arraybackslash}p{0.13\textwidth}}
\toprule
\textbf{Pathway} & \textbf{Potential system contribution} & \textbf{Critical boundary conditions} & \textbf{Key sources} \\
\midrule
On-site photovoltaic or wind with BESS & Reduce net import, absorb variable renewable output, provide resilience, and shift energy across hours. & Resource quality, hourly coincidence, storage duration, export limits, land, feeder hosting capacity, and battery lifecycle. & \cite{amin2023development,ljungblomoptimal,ref14,ref44} \\
Cogeneration, trigeneration, or fuel cells & Firm electricity and useful heat or cooling; possible reduction in grid import and local backup dependence. & Fuel carbon intensity, upstream emissions, heat-sink utilization, air quality, water, capacity factor, and islanding requirements. & \cite{ref212,ref14,siegle2025data} \\
Waste-heat recovery and thermal storage & Reduce cooling peaks, shift thermal demand, or supply nearby heat users. & Heat temperature, seasonal demand, distance to heat sink, piping, recovery energy, space, and thermal constraints. & \cite{ref14,gnibga2026improving,winick2025reducing} \\
Firm onsite supply and microgrids & Improve supply adequacy and resilience where renewable variability or network constraints are material. & Technology-specific licensing, lead time, project scale, cost, fuel or water requirements, islanding and protection, and shared-grid obligations. & \cite{ljungblomoptimal,ref212,ref14,ref264} \\
Temporal or geographic workload shifting & Align demand with renewable availability, lower congestion, and reduce marginal emissions when the destination and time are cleaner. & Service deadlines, data locality, network bandwidth, rebound, baseline, marginal rather than annual emissions, and destination capacity. & \cite{ref16,ref21,takci2025data,chen2026defer,ref219} \\
\bottomrule
\end{tabularx}
\end{table}

\subsection{Power quality design and asset monitoring}

Power quality mitigation should follow the measured or modeled emission spectrum and network impedance. Source-side converter design and filtering address harmonic emission \cite{9803460,9176824}. Phase allocation and neutral design address zero-sequence current \cite{gruzs1990survey,desmet2003analysis,mansoor1995predicting}. Transformer decisions require harmonic-loss and derating analysis \cite{bishop1996evaluating,faiz2015derating}, and cable decisions require ampacity analysis under nonsinusoidal current \cite{demoulias2007ampacity,tofoli2006study}. Resonance screening requires both source and network impedance \cite{sun2022data,sun2019modeling}. IEEE 519-2022 provides PCC distortion goals but does not select a facility-specific filter, transformer, conductor, or resonance remedy \cite{ieee519_2022}; those decisions depend on the measured spectrum, impedance, operating mode, and asset design.

Commissioning and periodic reassessment matter because server power supplies, rack density, cooling drives, UPS modes, filters, and firmware can change after energization. Temperature estimation, harmonic trending, loading history, and event-triggered waveform capture support asset decisions and model updates \cite{nair2026predictive,portillo2026monitoring,culler2026digital}. Table~\ref{tab:impact-solution} relates each impact to mitigation and present implementation maturity.

\begin{table}[!t]
\centering
\tabfont
\setlength{\tabcolsep}{4pt}
\renewcommand{\arraystretch}{1.1}
\caption{Impact-to-solution mapping and implementation maturity for data center grid integration}
\label{tab:impact-solution}
\begin{tabularx}{\textwidth}{>{\raggedright\arraybackslash}p{0.16\textwidth} Y Y >{\raggedright\arraybackslash}p{0.13\textwidth}}
\toprule
\textbf{Impact} & \textbf{Candidate mitigation} & \textbf{Implementation maturity and condition} & \textbf{Key sources} \\
\midrule
Interconnection delay and upgrades & Standard data, staged or flexible service, model validation, and cost-allocation clarity. & Official guidance and emerging tariff practice; outcomes remain jurisdiction-specific. & \cite{quint2025practical,force2025characteristics,ref246,ref259} \\
Distribution constraints & Upgrades, voltage control, non-wires alternatives, and DERMS or ADMS coordination. & Strong generic distribution methods; limited production evidence for hyperscale data center control. & \cite{fan2025reliability,castaneda2022electric,strezoski2023distributed,krishnamoorthy2026role} \\
Ride-through and reconnection & UPS or BESS coordination, staged recovery, disturbance records, and validated settings. & Modeling and emerging demonstration evidence; common-mode behavior needs system testing. & \cite{xie2025enhancing,alkhudaydi2026safe,conto2025texas,tao2025fast,ref248} \\
Harmonics and asset stress & Spectrum and impedance studies, filtering, phase balancing, thermal derating, and monitoring. & Established physical mechanisms; direct hyperscale asset-aging data remain limited. & \cite{9803460,sun2022data,demoulias2007ampacity,bishop1996evaluating,faiz2015derating} \\
Peak and market stress & Distinct products for shifting, capping, storage response, and firm curtailment. & Extensive optimization evidence; fewer verified field commitments. & \cite{ref16,alapera2018data,takci2025data,chen2026defer,ref267} \\
Renewable integration and supply adequacy & Co-location, storage, clean firm supply, microgrids, and carbon-aware scheduling. & Technology-specific evidence; must include deliverability, backup, lifecycle, and shared-grid obligations. & \cite{amin2023development,mohammadi2026grid,azizi2026strengthening,ref212,ref264} \\
\bottomrule
\end{tabularx}
\end{table}

\section{Research gaps and priorities}
\label{sec:research-agenda}

Nine connected research priorities emerge from the technical and implementation gaps identified above.

\begin{table}[!tp]
\centering
\tabfont
\setlength{\tabcolsep}{4pt}
\renewcommand{\arraystretch}{1.0}
\caption{Research gaps and suggested technical outputs}
\label{tab:research-gaps}
\begin{tabularx}{\textwidth}{>{\raggedright\arraybackslash}p{0.16\textwidth} Y Y >{\raggedright\arraybackslash}p{0.13\textwidth}}
\toprule
\textbf{Gap} & \textbf{Current limitation} & \textbf{Suggested output} & \textbf{Key sources} \\
\midrule
POI measurements & Limited public data across operating and disturbance regimes. & Anonymized waveform, phasor, spectrum, telemetry, and event data sets with metadata. & \cite{chatterjee2026measurement,portillo2026monitoring,culler2026digital,ref248} \\
Distribution performance requirements & Bulk-system guidance does not fully define feeder and asset criteria. & Risk-tiered utility rules for power quality, ride-through, recovery, thermal duty, and telemetry. & \cite{fan2025reliability,mitra2023analyzing,ref244,ref248} \\
Multi-fidelity models & Static, dynamic, impedance, and EMT studies often use disconnected assumptions. & Versioned model suite with shared validation cases and sensitivity-tested boundaries. & \cite{osti_3013288,lu2026dynamic,talukdar2026modeling,li2025ai,reyes2024effects} \\
Asset health integration & Transferable thermal methods exist, but direct data center asset data are sparse. & Spectrum-informed transformer, cable, and neutral models linked to planning decisions. & \cite{nair2026predictive,demoulias2007ampacity,bishop1996evaluating,faiz2015derating,cazacu2018thermal} \\
Flexibility valuation & Modeled technical potential is often not separated from verified availability. & Product-specific baselines, service constraints, rebound tests, and measurement protocols. & \cite{alapera2018data,takci2025data,chen2026defer,tao2025fast,ref219} \\
Sustainability counterfactuals & Annual matching can omit curtailment, marginal emissions, fuel and heat-sink effects, network duty, and lifecycle impacts. & Comparative co-location and scheduling studies with grid, carbon, water, land, heat, and asset metrics. & \cite{ref14,mohammadi2026grid,ljungblomoptimal,siegle2025data,ref212} \\
Renewable-rich feeder validation & Optimization and generic DER studies rarely measure co-located generation, computing, storage, reverse flow, power quality, and asset duty together. & Multi-season campus demonstrations with synchronized electrical, computing, thermal, and asset health data. & \cite{castaneda2022electric,padullaparti2022evaluation,strezoski2023distributed,amin2023development} \\
International comparative evidence & Current implementation evidence is concentrated in North America and selected European cases. & Common technical and sustainability metrics applied across regulatory, market, and grid contexts. & \cite{ref3,ljungblomoptimal,takci2025data,ref259,ref264} \\
Tariff and governance linkage & Technical capability is not always linked to enforceable service and update obligations. & Tariffs and agreements tied to maturity, models, telemetry, curtailment, recovery, and cost allocation. & \cite{anderson2025policy,ref259,ref263,ref266} \\
Cyber-physical resilience & Programmable load behavior is rarely tested jointly with grid dynamics and protection. & Integrated attack, detection, fail-safe, and recovery benchmarks. & \cite{shamseldein2026deep,huang2026cascading,barati2025resilience} \\
\bottomrule
\end{tabularx}
\end{table}

\begin{enumerate}
\item \textbf{Point of interconnection measurements.} Public or anonymized data sets should include active and reactive power, voltage and current waveforms, spectra, frequency response, protection or UPS states, recovery records, and the metadata needed for model validation. Existing measurement studies provide a starting point, but most models still rely on laboratory, simulation, or limited facility data \cite{chatterjee2026measurement,portillo2026monitoring,culler2026digital}.

\item \textbf{Distribution performance requirements.} Bulk-system large-load guidance should be complemented by risk-based utility criteria for harmonics, voltage sensitivity, ride-through, reconnection, thermal loading, telemetry, and protection at the service voltage \cite{fan2025reliability,mitra2023analyzing,ref248}.

\item \textbf{Updatable multi-fidelity models.} Static screening, phasor-domain dynamics, calibrated aggregates, impedance representations, and EMT submodels should share validation cases and version governance. The desired output is a study-oriented model portfolio, not one universal equivalent \cite{osti_3013288,lu2026dynamic,talukdar2026modeling}.

\item \textbf{Asset health in planning.} Interconnection and expansion tools should translate measured spectra and duty cycles into transformer loss and aging \cite{mitra2023analyzing,bishop1996evaluating,faiz2015derating}, cable ampacity \cite{demoulias2007ampacity,elsharkawy2022effect}, and neutral heating \cite{desmet2003analysis} before setting upgrade timing and flexible operating limits.

\item \textbf{Service-specific flexibility valuation.} Studies should report availability, baseline, response time, duration, rebound, state-of-charge or thermal constraints, service cost, and measurement and verification separately for temporal shifting, geographic shifting, power capping, storage response, and firm curtailment \cite{takci2025data,chen2026defer,tao2025fast}.

\item \textbf{Sustainability counterfactuals.} Co-location and flexible operation should be evaluated against explicit alternatives using renewable curtailment, marginal emissions, water and land use where relevant, lifecycle effects, network upgrades, shared-grid services, and asset life. Cogeneration analyses need fuel and heat-sink carbon accounting; renewable co-location analyses need time-varying import and export envelopes \cite{ref14,ljungblomoptimal,siegle2025data}.

\item \textbf{Renewable-rich campus validation.} Field studies should measure renewable output, computing demand, storage, voltage-control action, reverse flow, harmonics, transformer and cable duty, and curtailment at the same site. Such measurements would test whether coordination increases hosting capacity without transferring stress to other assets \cite{castaneda2022electric,padullaparti2022evaluation,amin2023development}.

\item \textbf{Comparative regulation and tariffs.} Research should connect payment and service terms to verified technical performance and compare project-maturity tests, cost allocation, flexible service, co-location, forecasting, model updates, recovery duties, and behind-the-meter supply across diverse grid, market, and regulatory contexts. International energy-system context is available \cite{ref3,ljungblomoptimal,takci2025data}, whereas the current North American implementation record is represented by large-load policy and FERC proceedings \cite{anderson2025policy,ref259,ref264}.

\item \textbf{Cyber-physical resilience.} Computational loads should be studied as programmable grid interfaces because beneficial control and malicious or correlated modulation may use the same actuation paths. Benchmarks should combine cyber scenarios with converter, protection, workload, and system-dynamic models and should test detection, fail-safe behavior, and recovery \cite{shamseldein2026deep,huang2026cascading,barati2025resilience}.
\end{enumerate}
Table~\ref{tab:research-gaps} links these priorities to specific technical outputs.

\section{Conclusion}
\label{sec:conclusion}

Data center grid impacts are governed by more than annual electricity use or requested peak demand. Siting, project maturity, coincidence with system peaks, ramping, converter and UPS behavior, voltage sensitivity, ride-through, recovery, harmonic spectra, and verified flexibility determine the consequences for generation adequacy, transmission planning, distribution operation, and markets. Interconnection studies should therefore be based on staged and updatable facility data rather than a single static megawatt value.

Distribution-level mechanisms are technically credible but remain less well documented in hyperscale field settings than bulk-system impacts, even though a growing share of new large-load requests is arriving at distribution utilities. Harmonic and sustained current can increase transformer and cable thermal duty, triplen components can load neutral conductors, frequency-dependent impedance can produce resonance, and control or protection states can shape disturbance response. Case-specific measurement, commissioning, spectrum and impedance analysis, thermal assessment, and model revalidation are more defensible than universal derating or protection assumptions.

Sustainable integration requires a similarly conditional assessment. Data centers may absorb variable renewable output, shift computing, provide storage-mediated response, or support clean supply development, but they may also add peak demand, backup obligations, reverse power flow, rebound, water use, and asset stress. Annual renewable-energy matching or physical co-location alone does not establish a net system benefit. Credible claims should account for hourly and locational supply, network deliverability, marginal emissions, lifecycle effects, water and land impacts, shared-grid obligations, and the operating capability that is actually measured or contractually available. Coordinated planning, validated models, transparent tariffs, interoperable controls, and asset-health monitoring are therefore central to the reliable and sustainable growth of data center infrastructure in the grid.

\section*{CRediT authorship contribution statement}
\textbf{Bibek Dhakal:} Conceptualization, Methodology, Investigation, Formal analysis, Data curation, Visualization, Writing -- original draft, Writing -- review \& editing.
\textbf{Manohar Chamana:} Conceptualization, Supervision, Project administration, Funding acquisition, Writing -- review \& editing.
\textbf{Biswambhar Gayen:} Resources, Writing -- original draft.
\textbf{Jatendar Kumar:} Writing -- original draft.
\textbf{S M Shams Rahman:} Visualization.
\textbf{Aaqib Peerzada:} Writing -- original draft, Writing -- review \& editing.
\textbf{Ki Yeob Lee:} Writing -- original draft, Writing -- review \& editing.
\textbf{Olatunji M. Adeyanju:} Writing -- review \& editing.
\textbf{Bhaskar Mitra:} Supervision, Writing -- review \& editing.
\textbf{Stephen Bayne:} Funding acquisition, Supervision, Writing -- review \& editing.

\section*{Declaration of competing interest}
The authors declare that they have no known competing financial interests or personal relationships that could have appeared to influence the work reported in this paper.

\section*{Funding}
This material is based upon work supported by the U.S. Department of Energy, Office of Cybersecurity, Energy Security, and Emergency Response (CESER), under Award Number DE-CR0000038.

\section*{Acknowledgements}
The authors gratefully acknowledge support from the U.S. Department of Energy (DOE) and its Office of Cybersecurity, Energy Security, and Emergency Response (CESER). The pictograms in Figures~\ref{fig:grid-layer-taxonomy} to~\ref{fig:solution-architecture} are from the Solar Icon Set by 480 Design (CC BY 4.0, \url{https://creativecommons.org/licenses/by/4.0/}), obtained through SVG Repo and recolored by the authors, and from Microsoft Visio stencils. Pacific Northwest National Laboratory is operated for the U.S. Department of Energy by Battelle Memorial Institute under Contract DE-AC05-76RL01830.

\section*{Disclaimer}
This work was prepared as an account of work sponsored by an agency of the United States Government. Neither the United States Government nor any agency thereof, nor any of their employees, makes any warranty, express or implied, or assumes any legal liability or responsibility for the accuracy, completeness, or usefulness of any information, apparatus, product, or process disclosed, or represents that its use would not infringe privately owned rights. Reference herein to any specific commercial product, process, or service by trade name, trademark, manufacturer, or otherwise does not necessarily constitute or imply its endorsement, recommendation, or favoring by the United States Government or any agency thereof. The views and opinions expressed herein do not necessarily state or reflect those of the United States Government or any agency thereof.

\section*{Data availability}
No data were used for the research described in the article.

\begingroup
\sloppy
\bibliographystyle{elsarticle-num}
\bibliography{references}
\endgroup

\end{document}